\documentstyle[epsf]{mn}

\title{Cross-correlation of the unresolved X-ray background with faint galaxies}
\author[A. Newsam, I. M$^{\it c}$Hardy, L. Jones and K. Mason]{
  A.M.~Newsam,$^{1,2,}$\thanks{Email:~amn@astro.livjm.ac.uk}
  I.M.~M$^{\rm c}$Hardy,$^1$ L.R.~Jones,$^{3}$
        K.O.~Mason$^4$\\
  $^1$Department of Physics and Astronomy, University of Southampton, 
	Southampton, SO17 1BJ.\\
  $^2$Astrophysics Research Institute, Liverpool John Moores University,
	Liverpool, L3 3AF.\\
  $^3$School of Physics and Space Research, University of Birmingham,
	Edgbaston, Birmingham, B15 2TT.\\
  $^4$Mullard Space Science Laboratory, University College London, 
        Holmbury St Mary, Dorking, RH5 6NT.}
\date{}

\def\etal{\em et~al.}
\def\rosat{\sc Rosat}

\def\lxlr{${\rm L}_{\rm X}/{\rm L}_{\rm R}$}
\def\lglxlopt{$\log[{\rm L}_{\rm X}/{\rm L}_{\rm opt}$]}

\def\scite#1{\def\citename##1##2{##1}\csname b@#1\endcsname
 \shortcite{#1}}
\def\pcite#1{\def\citename##1##2{##1##2}\csname b@#1\endcsname\nocite{#1}}
\def\ycite#1{\def\citename##1##2{##2}\csname b@#1\endcsname\nocite{#1}}
\def\ncite#1{\def\citename##1##2{##1}\csname b@#1\endcsname\nocite{#1}}

\begin{document}

\maketitle

\begin{abstract}

At the faint end of the deepest X-ray surveys, a population of X-ray
luminous galaxies is seen. In this paper, we present the results of a
cross-correlation between the residual, unresolved X-ray photons in a
very deep X-ray survey and the positions of faint galaxies, in order
to examine the importance of these objects at even fainter flux
levels. We measure a significant correlation on all angular scales up
to $\sim$1 arcmin. This signal could account for a significant
fraction of the unresolved X-ray background, approximately 35 per cent
if the clustering is similar to optically selected galaxies. However,
the angular form of the correlation is seen to be qualitatively
similar to that expected for clusters of galaxies and the X-ray
emission could be associated with hot gas in clusters or with QSOs
within galaxy clusters rather than emission from individual faint
galaxies. The relative contribution from each of these possibilities
cannot be determined with the current data.

\end{abstract}


\section{Introduction}

The nature of the sources and emission mechanisms that contribute to the
cosmic X-ray background (XRB) remains one of the major questions in
astrophysics. Deep surveys, particularly with the {\rosat} satellite, have
resolved a significant fraction ($\sim$50 per cent) of the XRB, with optical
identification of the sources enabling classification of much of the emission
(eg \pcite{mn_nov}, \pcite{Schmidt+98}), but a number of important questions
still remain. From current surveys, it is seen that at least 30 per cent of
the X-ray background can be attributed to broad-line QSOs, but the steep X-ray
spectra of QSOs does not match the shallow spectrum of the residual XRB.
Therefore, a population of faint X-ray sources with flatter spectra is
required to make up much of the remainder of the XRB.

At the faint end of the deepest surveys just such a population is emerging with
increasing numbers of X-ray luminous galaxies with narrow optical emission
lines (NELGs) (\pcite{mn_nov}, \pcite{Boyle+95}). However, current surveys are
only just beginning to see significant numbers of such objects at their
faintest limits and so the significance of these new objects to the XRB as a
whole is highly uncertain.

One can get deeper than the resolution limit of surveys by looking at
the correlation between the {\em unresolved\/} regions of a deep X-ray
survey and the positions of a population of putative X-ray sources, in
this case galaxies. \scite{Roche+95} use such a cross-correlation
method to show that faint galaxies are a significant contributor to
the unresolved flux in three deep {\rosat} PSPC observations (two of
$\sim$25 ksec, and one of $\sim$50 ksec). After the removal of
resolved sources, they find a highly significant detection
($\sim$5$\sigma$) in a correlation between the X-ray photons
(0.5~--~2.0 keV) and the positions of 18$\le$B$\le$23 mag galaxies. In
\scite{Roche+96} they repeat the calculation using a slightly deeper
X-ray observation (74 ksec). In their analysis, they apply the
formalism of \scite{Treyer+96} to model the clustering and evolution
of the population of X-ray sources, in an attempt to correct for
contamination to the cross-correlation signal due to clustering of the
sources. Although a significant signal is again seen, the
uncertainties in the assumptions required by the method mean that they
are unable to draw any firm quantitative conclusion about the
contribution to the XRB from faint galaxies.  Nevertheless, an
extrapolation of their results to high redshifts implies that
$\sim$~30{--}50 per cent of the total 0.7{--}2.0 keV X-ray background
might be due to emission from faint galaxies.

Another analysis by \scite{Almaini+97} using three PSPC observations (each
of~$\sim$50 ksec) also shows a significant signal. They again apply the
formalism of \scite{Treyer+96}, with modifications to compensate for the
point spread function of the PSPC, and an extrapolation to high redshift gives
a contribution to the XRB from faint galaxies of $\sim$ 40 $\pm$ 10
per~cent. However, they note that this estimate has a strong dependence on the
assumed evolution, distribution and clustering properties of the galaxies.

On a wider scale, but with shallower observations, \scite{Soltan+97} correlate
the positions of galaxies with the {\rosat} All-Sky Survey
\cite{Snowden+90} and find a similar signal to {\scite{Roche+96} and
{\scite{Almaini+97}.

In this paper we correlate the positions of faint galaxies with a very deep
PSPC X-ray observation (115 ksec). In this observation, a significant fraction
of the {\em resolved\/} X-ray background photons are directly associated with
galaxies (the NELGs~--~ see \pcite{mn_nov}). Therefore, not only will a
cross-correlation analysis enable us to probe further into the unresolved XRB
than previous studies, but it will provide a test of whether the contribution
to the XRB from galaxies extends to significantly fainter fluxes than the
limits of shallower surveys or whether they contribute only over a relatively
narrow range in flux. This will give us a clearer idea of the nature of the
contribution of NELG-like objects to the XRB.

In section~\ref{sec:data} we describe the X-ray and optical data used in this
study and give details of the cross-correlation method employed. We also
highlight some of the problems associated with attaching a significance to the
results and describe the simulations we have used to determine accurate error
estimates. In section~\ref{sec:results} we present the results of applying the
cross-correlation using galaxies from a selection of magnitude ranges. The
possible implications of these results are discussed in
section~\ref{sec:discussion} and we present our conclusions in
section~\ref{sec:conclusions}.

\section{Data and analysis}\label{sec:data}

The X-ray data used in this analysis come from the UK {\rosat} Deep Survey
described in detail in \scite{mn_nov} and \scite{GBR}. The data consist of a
total of 115 ksec of {\rosat} position sensitive proportional counter (PSPC)
observations of RA~$13\,\, 34\,\, 37.0$ Dec~$+37\,\, 54\,\, 44$ (J2000), a
region of sky selected because of its extremely low obscuration~---~$N_H \sim
6.5 \times 10^{19} \hbox{cm}^{-2}$. Only the inner 15 arcmin radius of the
PSPC field of view is used in this study, where sources have been detected
and, in many cases, optically identified down to a flux limit of $2 \times
10^{-15} \hbox{ erg cm$^{-2}$ s$^{-1}$}$ (0.5-2 keV~--~all fluxes in this
paper will refer to this band), resolving approximately 50 per cent of the
cosmic X-ray background (XRB).

In this analysis, we wish to study the unresolved component, so these sources
must be ``masked out''. Because of the large range of brightnesses in the
survey (up to $4.8 \times 10^{-13} \hbox{ erg cm$^{-2}$ s$^{-1}$}$), and the
variation of the PSPC point spread function over the image, a fixed mask size
is inappropriate. We therefore use a gaussian approximation to the PSPC point
spread function from \scite{HasingerPSF} to select a mask radius for each
source that leaves a residual of $\sim$0.1 photons, assuming that it is a
point source.  For an on-axis source at the detection limit, this gives a
mask radius of 29 arcsec and excludes 99.5 per cent of the source photons.

The galaxy identifications are taken from deep R-band CCD imaging of the
survey region, using the University of Hawaii 8K$\times$8K CCD array
\cite{8kpaper} on CFHT with a 1 hour exposure, giving galaxies to
R$\sim$24.5. Objects were found using the PISA software provided by
Starlink. Galaxies were separated from stars using the ratio between the
aperture magnitude of each object to its peak count in any one pixel. Stellar
objects have an approximately constant ratio with the more diffuse galaxies
forming a distinct population. Plots of peak counts against magnitude can,
therefore, be used to separate galaxies from stars as in
figure~\ref{fig:starsep}.
\begin{figure}
 \epsfxsize 0.95\hsize
 \epsffile{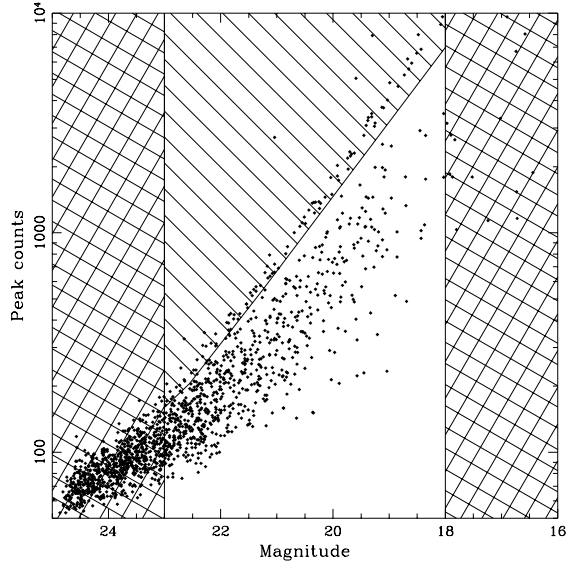}
 \caption{
The galaxy selection criteria for the objects found on the R-band CCD
images. The points are the objects detected by the PISA software system, the
cross-hatched regions show those areas excluded by the upper and lower
magnitude limits and the diagonally-hatched region indicates those objects
rejected because they are point-like. The objects well above the line of
stellar objects are cosmic ray defects in the images. All objects outside the
hatched regions are included in our analysis. Only a random sub-sample of the
detected objects has been shown here for clarity.
}\label{fig:starsep}
\end{figure}

Regions of the image contaminated by bright stars are excluded. The total
area of overlap between residual (ie unmasked) XRB and useful R-band image is
0.052 degree$^2$, approximately 26 per cent of the 15 arcmin radius region of
the Deep Survey image. For this study, we use galaxies with 18$<$R$\le$23. At
fainter levels, the separation between point-like and extended objects
becomes uncertain (see figure~\ref{fig:starsep}), and at brighter levels, the
number density of galaxies becomes small and field-to-field fluctuations
would dominate any conclusions about the XRB as a whole. In total, the
overlap region contains 1451 galaxies within this magnitude range.

The cross-correlation method is similar to that of \scite{Roche+95}. The
number of X-ray photons per pixel (2 arcsec square) in a series of annuli
from $\theta$ to $\theta + \Delta\theta$ around each galaxy is obtained and
the number expected from a random distribution normalised to the mean
intensity of the masked image is subtracted. The contribution from all the
galaxies is then averaged:
\begin{equation}
 {\cal X}_{\rm xg}(\theta) = 
    { { \sum_{N_{gal}}\; (N_x(\theta) - N_p(\theta) \overline{N_x}) }
	\over
      { N_{gal} \; A(\theta) } }
\label{eqn:xgg}\end{equation}}%
\noindent
where ${\cal X}_{\rm xg}(\theta)$ is the X-ray photon/galaxy
cross-correlation signal for aperture }$\theta$ (in photons galaxy$^{-1}$
arcsec$^{-2}$), $N_{gal}$ is the number of galaxies in the overlap region,
$N_x(\theta)$ is the number of X-ray photons within the aperture around a
particular galaxy, $N_p(\theta)$ is the number of pixels in the aperture,
$\overline{N_x}$ is the average number of X-ray photons per pixel and
$A(\theta)$ is the area of the aperture in arcseconds.

It should be noted that for larger annuli, the area of the X-ray image
covered, and hence the number of galaxies that contribute to the
cross-correlation, is slightly larger than the values given above.

\subsection{Error estimation}

\scite{Roche+95} estimate errors on ${\cal X}_{\rm xg}(\theta)$ using a 
bootstrap technique, but this does not take into account the problems
associated with spurious apparent correlations produced by the
auto-correlation functions of the distribution of unmasked X-ray photons and
the regions excluded on the R-band CCD image. We have, therefore, performed a
series of Monte Carlo simulations to estimate the significance of our
results.

Two sets of simulations were performed, both using the actual, masked
distribution of X-ray photons, but randomising the distribution of galaxies
in different ways. In both cases those regions of the CCD image excluded were
matched to the actual data.

For the first set of simulations, galaxy positions were chosen entirely at
random until the observed number of galaxies were obtained. However, although
these simulations will include the effects of the CCD selection, and X-ray
source masking and photon auto-correlation, they will not include any effect
from the galaxy-galaxy angular correlation function. In order to estimate
whether this effect is significant, we performed a further series of
simulations. In these, instead of entirely random galaxy positions, we divided
the actual galaxy distribution into a set of 54$\times$66 arcsec ``boxes'' and
shuffled these boxes around at random.

\begin{figure}
 \epsfxsize 0.95\hsize
 \epsffile{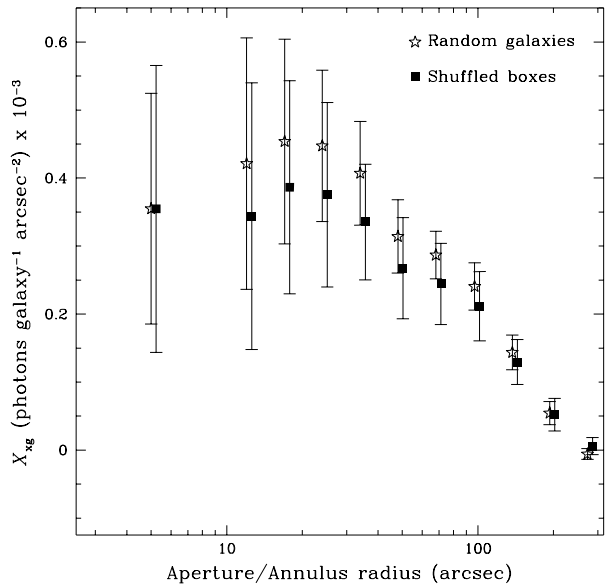}
 \caption{
A comparison of the two different methods of Monte Carlos simulations
used. The stars show the means of 100 simulations using randomly positioned
galaxies, and the filled boxes the means for a similar number of simulations
where ``boxes'' of observed galaxy positions have been shuffled (see
text). The error-bars show the 1-$\sigma$ scatter in the simulations and the
``shuffled'' points have been moved slightly to the right for clarity.
}\label{fig:MCcomp}
\end{figure}

A comparison of these two methods is shown in figure~\ref{fig:MCcomp}. There is
no significant difference between the two, indicating that the effect of the
galaxy-galaxy correlation function on these scales is negligible. For the rest
of the paper, we will only consider the results from the first set of Monte
Carlos (ie those with entirely random galaxy positions).

\section{Results}\label{sec:results}

\begin{figure}
 \epsfxsize 0.95\hsize
 \epsffile{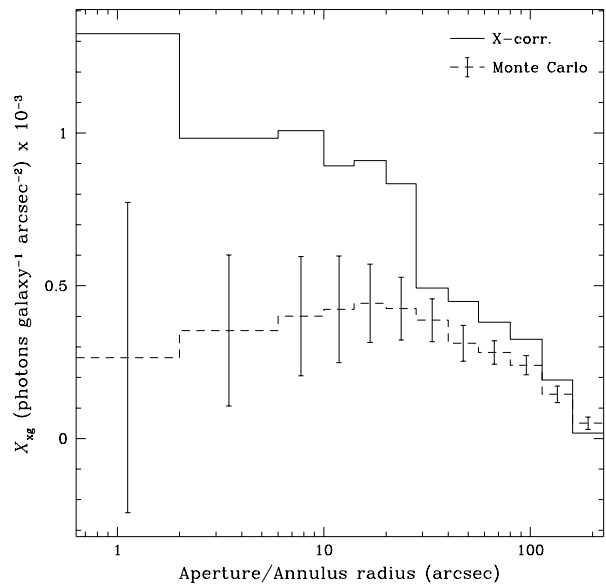}
 \caption{
The cross-correlation of 18$<$R$\le$23 mag galaxies with the unresolved 0.5-2keV
X-ray background in a series of annuli. The solid line shows the actual
cross-correlation and the dashed line, the mean result of a series of
simulations using the actual residual X-ray background image but a random
distribution of galaxies. The error-bars show the 1-$\sigma$ scatter of the
simulations about the mean.}\label{fig:18-23xcor}
\end{figure}
The cross-correlation signal seen for all galaxies with 18$<$R$\le$23 is given
in figure~\ref{fig:18-23xcor} together with the mean and 1-$\sigma$ scatter of
the Monte Carlo simulations. There is a significant correlation above that
expected from a random distribution of galaxies out to a radius of $\ga$ 1
arcmin.

We can estimate the fraction of the unresolved XRB in this field associated
with galaxies by taking a 1~arcmin radius aperture around each galaxy and
summing up the total number of X-ray photons detected in each aperture above
that expected from a random distribution. The random expectation in each
aperture will be affected by the masking of both the X-ray and optical images
and so was determined from a set of simulated PSPC images. Once an initial
estimate of the contribution was determined, the process was repeated but
this time the simulated X-ray images were created with a corresponding
fraction of the X-rays associated with galaxies. This process was iterated
until convergence was reached. The scatter in the counts for the final
simulations was then used to estimate an error on the contribution to the
residual XRB. We find that 67~$\pm$~9 per cent of the unresolved XRB photons
are associated with galaxies.

However, this result does not take into account any clustering of galaxies on
scales up to 1 arcmin. If such clustering is present, this photon excess will
be an over-estimate since each galaxy will produce a correlation with the
X-ray emission of its clustered companions. We can approximately correct for
this, following the procedure of \scite{Roche+95}, by dividing the excess
number of galaxy-photon pairs by $1 + N_{gg}'(\theta < 1{\hbox{~arcmin}}) /
N_{gal}$ where $N_{gg}'(\theta < 1{\hbox{~arcmin}})$ is the excess number of
galaxy-galaxy pairs with separation less than 1 arcmin and $N_{gal}$ is the
total number of galaxies. Applying this correction reduces our result to
35~$\pm$~5 per cent of the unresolved XRB associated with galaxies (where the
error is only from the scatter in the simulations). However, it is important
to realise that this is only very approximate and, in particular, is based on
the {\em average\/} galaxy-galaxy correlation. If X-ray emission is
preferentially associated (or disassociated) with clustered regions, this
correction will be an under- (over-) estimate. We will return to this
question later.

As discussed by \scite{Roche+95}, there are two further effects which may
lead to overestimation of the contribution: (i) X-ray emission from galaxies
with R$>$23 clustered with the R$\le$23 galaxies and (ii) correlation from
galaxies clustered with X-ray emitting QSOs. Both of these effects are very
difficult to quantify. An R$=$23 galaxy just below the detection limit of the
survey ($2 \times 10^{-15} \hbox{ erg cm$^{-2}$ s$^{-1}$}$) would have an
X-ray to R-band luminosity ratio of {\lxlr}$\sim$0.8, which is consistent
with the NELGs resolved in the survey which have $0.003 \la$ {\lxlr} $\la
1.5$. However, the numbers of such objects, and the extent to which they
cluster with brighter galaxies is not known. \scite{Roche+96} and
\scite{Almaini+97} attempt to account for (i) and the clustering of observed
galaxies using a formalism developed by \scite{Treyer+96}. However, this
method, which models the evolution and clustering of the X-ray sources, is
sensitive to a number of assumptions. In particular, it is assumed that the
galaxies are all drawn from a single population of X-ray sources with $L_{\rm
X} \propto L_{\rm opt}$ for all galaxies at all redshifts. However, here we
would expect our catalogue of galaxies to contain a combination of ``normal''
galaxies ({\lglxlopt}$\; \la -2$), NELGs ({\lglxlopt}$\; \la 1$) and clusters
of galaxies ({\lglxlopt}$\; \la 1.5$) (eg \pcite{mn_nov} and
\pcite{Stocke+91}). In addition, variations in the models for the clustering
and evolution of the galaxies can add large uncertainties to the formalism
\cite{Almaini+97}.

It has been seen (eg \pcite{Smith+95}) that galaxy clustering around X-ray
selected AGN is similar to that of galaxy-galaxy clustering and the X-ray
emissivity estimated by \scite{Roche+95} from their correlation is larger
than that found for the local AGN emissivity \cite{Miyaji+94}, and so they
choose to neglect possible contamination from AGN associated with clustered
galaxies. However, from these arguments alone, it is not possible to exclude
a significant fraction of the observed correlation being due to this effect,
particularly since the emissivity must be calculated {\em assuming\/} that
the effect is negligible.

We have, therefore, repeated the correlation analysis with the galaxies
divided into ``bright'' ($18 < {\rm R} \le 22$~---~686 galaxies) and
``faint'' ($22 < {\rm R} \le 23$~---~765 galaxies) populations thereby
probing different redshift distributions.
\begin{figure}
 \epsfxsize 0.95\hsize
 \epsffile{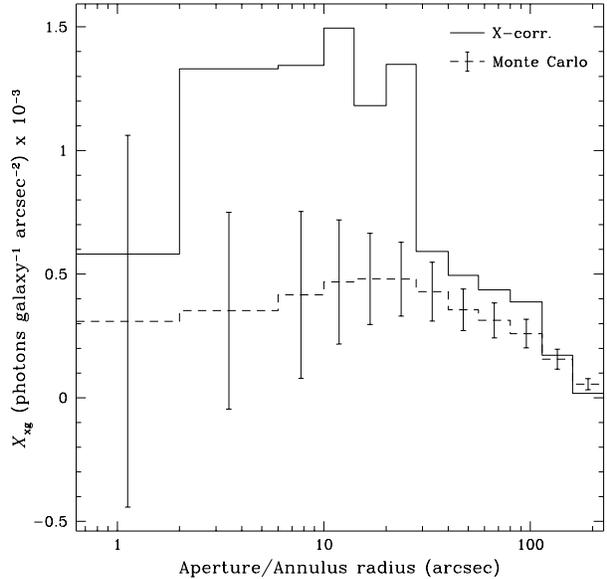}
 \caption{
As figure~\ref{fig:18-23xcor} but for galaxies with 18$<$R$\le$22.
}\label{fig:18-22xcor}
\end{figure}
\begin{figure}
 \epsfxsize 0.95\hsize
 \epsffile{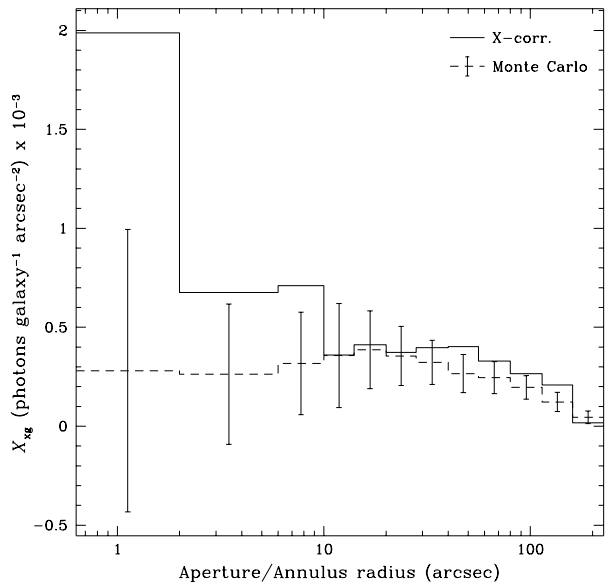}
 \caption{
As figure~\ref{fig:18-23xcor} but for galaxies with 22$<$R$\le$23.
}\label{fig:22-23xcor}
\end{figure}
The results are shown in figures~\ref{fig:18-22xcor} and~\ref{fig:22-23xcor}
for ``bright'' and ``faint'' galaxies respectively.  The angular distribution
of the correlation signal is clearly very different in the two cases with the
correlation from the fainter galaxies being dominated by annuli of $<$10
arcsec, and the brighter galaxies contributing on larger angular scales. We
will return to this in section~\ref{sec:discussion}. It is important to
realise that these two measurements, although based on distinct populations
of galaxies, are not independent.  Clearly, any contribution from clusters of
galaxies and galaxies associated with QSOs will affect both correlations and
the ``bright'' correlation will contain a signal from clustering around X-ray
emitting galaxies with R$>$22.

\section{Discussion}\label{sec:discussion}

We can get some idea of the possible contribution to the correlation
from QSOs by extrapolating the distribution of identified sources in
the Deep Survey below the flux limit. Using fits to the source counts
as a function of flux given in \scite{mn_nov} we find that a simple
extrapolation would resolve the entire XRB at a flux of $\sim 1 \times
10^{-17} \hbox{ erg cm$^{-2}$ s$^{-1}$}$. Extrapolating the QSO fit
down to this limit gives an additional contribution to the {\em
unresolved\/} XRB of 6 per cent. However, the fit is not well
constrained and extrapolating the 1-$\sigma$ upper confidence limit to
the fit gives 37 per cent. Obviously, extrapolation of a simple linear
fit over such a large flux range is somewhat unreliable, but given the
increasing significance of NELGs at fainter fluxes in identified
surveys, it is clear that broad-lined QSOs are not likely to be the
sole, or maybe not even dominant contributor to the unresolved XRB. Of
course, this does not rule out an AGN-like origin for the X-rays since
any X-ray galaxies contributing to the correlation signal could be
low-luminosity or obscured AGN.

Nevertheless, the angular distribution of the ``bright'' and ``faint''
correlation signals may indicate a clustered environment for a large
fraction of the X-ray emitting objects. We can exclude the possibility
that the correlation at larger angles is due to the correlation with
residual photons in the wings of the ``masked'' sources since this
should account for less than 1 per cent of the residual XRB
photons. However, a handful of the known X-ray objects are identified
with small clusters of galaxies \cite{mn_nov} and these will be
slightly extended.  We therefore repeated the correlation using larger
masks around each source (sufficient to mask out all but 0.01 photons
from a point source) but observed no significant difference in the
correlation.

\subsection{The angular form of the cross-correlation}\label{sec:angular}

\begin{figure}
 \epsfxsize 0.95\hsize
 \epsffile{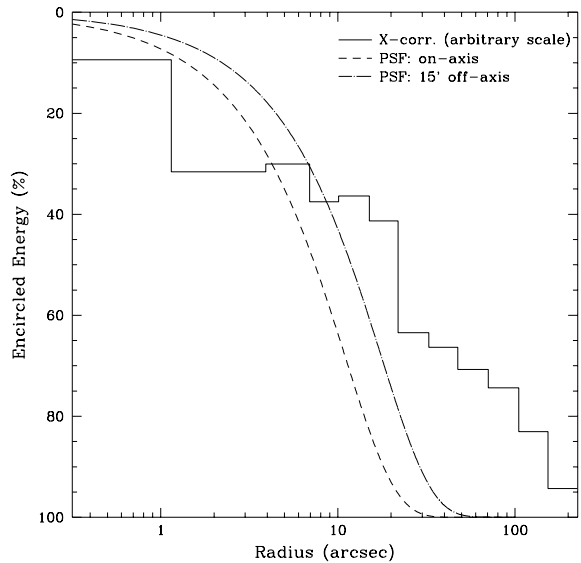}
 \caption{
The encircled energy as a function of angular radius for the PSPC point spread
function (PSF). The two dashed lines are the encircled energies of a gaussian
model of the PSF from \protect\scite{HasingerPSF} for an on-axis point source
(dashed line) and a point source 15 arcmin off-axis (dot-dash). Both are
evaluated at an energy of 1keV. The cross-correlation from
figure~\ref{fig:18-23xcor} is shown for comparison (solid line~---~arbitrary
scale).
}\label{fig:PSPC_PSF}
\end{figure}

The angular form of the cross-correlation signals that we see will depend on
both the point spread function (PSF) of the PSPC instrument (see
figure~\ref{fig:PSPC_PSF}) and the angular correlation between galaxies and
X-ray sources (whether galaxies themselves, QSOs or emission from hot gas in
galaxy clusters).

\begin{figure}
 \epsfxsize 0.95\hsize
 \epsffile{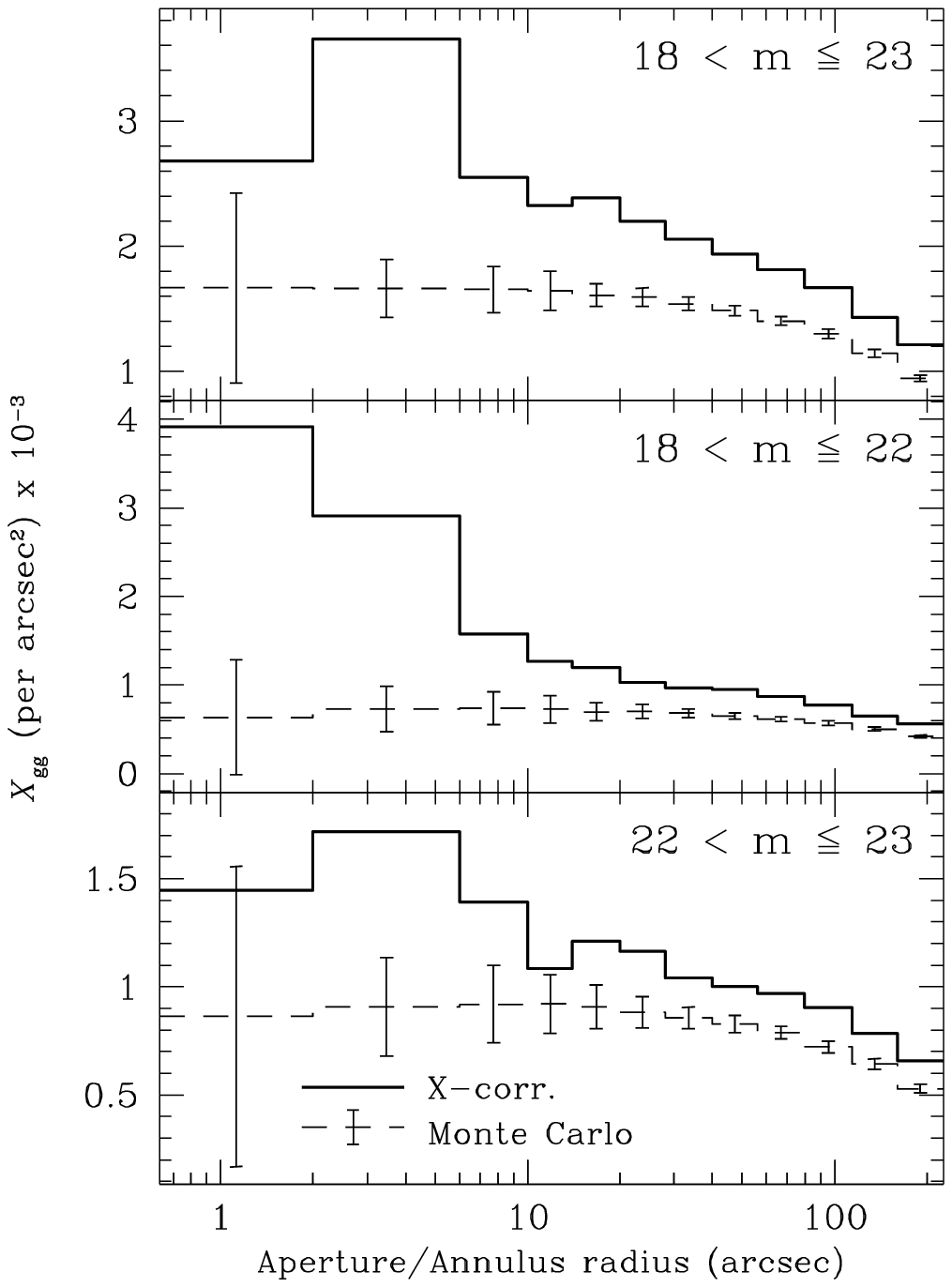}
 \caption{
The galaxy-galaxy cross-correlation for galaxies with 18$<$R$\le$23 (top),
18$<$R$\le$22 (middle) and 22$<$R$\le$23 (bottom). In each case, the solid line
shows the measured signal and the dashed lines and error bars, the average and
1-$\sigma$ scatter of a set of simulations using randomly selected galaxy
positions but the CCD masking and edge-effects of the actual data. A strong
positive signal is seen above the random expectation in each case.
}\label{fig:GGcor}
\end{figure}

Unfortunately, we cannot measure this angular correlation since we do not
know which are the X-ray sources. However, we can measure the overall
galaxy-galaxy correlation (${\cal X}_{\rm gg}$) for the different magnitude
ranges as shown in figure~\ref{fig:GGcor}.
These results show a qualitatively similar form to those of the
X-ray/galaxy cross-correlations (figures~\ref{fig:18-23xcor}
to~\ref{fig:22-23xcor}) but here, the ${\cal X}_{\rm gg}$ peak is at
small radii is for the ``bright'' galaxies, with a broader
distribution for the ``faint'' galaxies. Clearly, the correlation
between galaxies and X-ray sources can only be crudely approximated by
the overall galaxy-galaxy correlation.

However, the angular form of the X-ray/galaxy correlation signal that we see
is not well described by correlation with {\em unclustered\/} galaxies. This
can be seen from simulations where a given fraction of the residual XRB is
associated with randomly distributed galaxies. We have created a number of
these simulations for a range of XRB contributions. The distribution of
fluxes of X-ray sources is taken from an extrapolation of the source counts
in \scite{mn_nov} and sufficient of these sources are associated with
randomly distributed galaxies to produce a known fraction of the residual
XRB. X-ray sources with fluxes above the limit of the Deep Survey data ($2
\times 10^{-15} \hbox{ erg cm$^{-2}$ s$^{-1}$}$) are masked out in the same
way as the real data and the CCD masking and edge-effects are reproduced.

\begin{figure*}
 \epsfxsize 0.95\hsize
 \epsffile{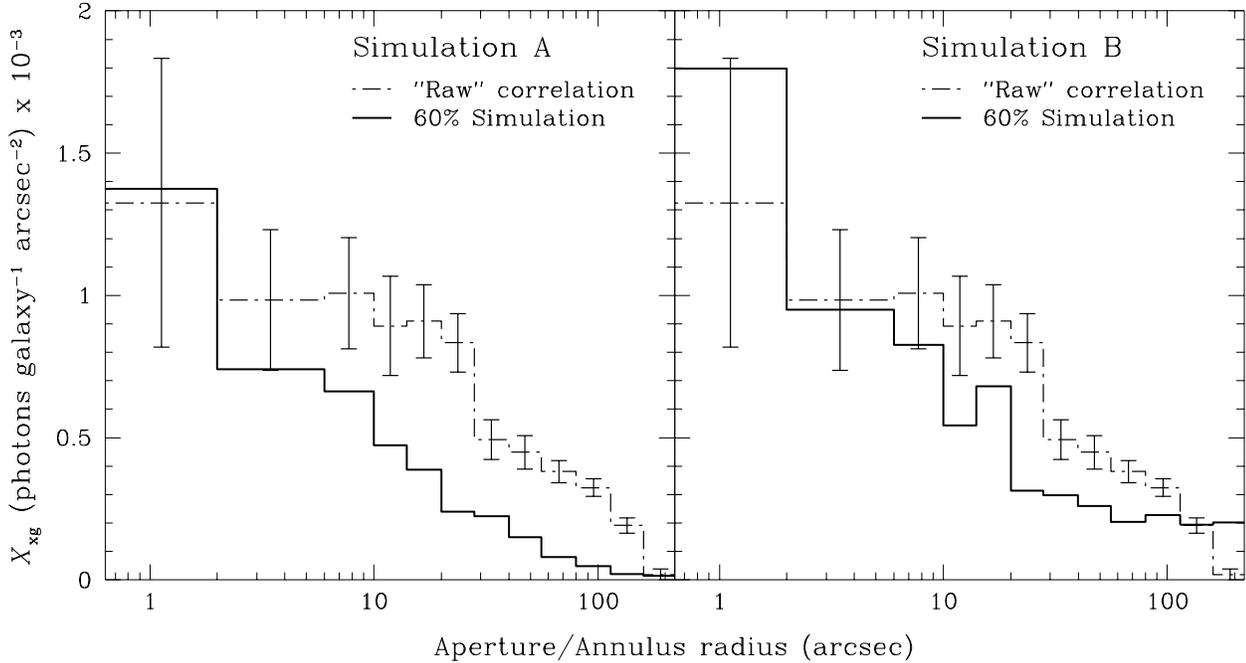}
 \caption{
Two realisations of simulated cross-correlations with 60 per cent of the
residual X-ray photons associated with (random) galaxies. Each panel shows
the results of one simulation. The ${\cal X}_{\rm xg}$ signal for the
simulated data set is shown with the solid line, and the dot-dashed line shows
the observed correlation from figure~\ref{fig:18-23xcor} for comparison. The
error bars are those from the Monte Carlo simulations also shown in
figure~\ref{fig:18-23xcor} and give an estimate of the errors on each
aperture, although they should be considered indicative only.
}\label{fig:simulat}
\end{figure*}

Results for two typical simulations are given in figure~\ref{fig:simulat}.
In both cases, 60 per cent of the unresolved XRB was associated with galaxies
(approximately matching the uncorrected value calculated in
section~\ref{sec:results}). Although the variation between simulations is
large, in all cases the simulation is more ``peaked'' than the observed
correlation~---~ie it has a higher fraction of its correlation in small
apertures. To quantify this, we define the statistic ${\cal P}:$
\begin{equation}
  {\cal P} = { {X(0, 10 \hbox{~arcsec})} 
	\over 
	{X(10 \hbox{~arcsec},  1 \hbox{~arcmin}) } }
\end{equation}
where $X(\theta_1, \theta_2)$ is the sum of the apertures with radii
$\theta_1$$<$$\theta$$\le$$\theta_2$. For the observed correlation we find
${\cal P}_{\hbox{\footnotesize obs}} = 0.93$ whereas for 50 simulations, with
an imposed residual XRB contribution from galaxies of 60 per cent, we find
${\cal P}_{\hbox{\footnotesize sim}} = 1.91 \pm 0.35$ where the error is the
1-$\sigma$ scatter of the simulations. Clustering, therefore, clearly plays a
role in the angular form of the signal that we see.

\subsubsection{X-ray emission from clusters of galaxies}

One obvious possibility for some of the X-ray/galaxy cross-correlation
signal is emission from the hot gas in the intra-cluster medium of
galaxy clusters or groups. If we assume that the X-ray emission from
such a cluster at a moderate redshift (eg $z \ga 0.3$) is well
approximated by a point source in the PSPC, we would expect the form
of the cross correlation in a series of annuli of inner and outer
radii $\theta_1$ and $\theta_2$ respectively, $W(\theta_1, \theta_2)$,
to be approximately described by:
\begin{equation}
W(\theta_1, \theta_2) \approx 
    {  {N_g(\theta_1, \theta_2)} \over {N_g(0, \theta_A)} }
    \times
    {  {P(\theta_1, \theta_2)} \over {P(0, \infty)} }
\label{eqn:clustprof}\end{equation}
where $N_g(\theta_1, \theta_2)$ is the number of galaxies expected
from the cluster in the annulus between $\theta_1$ and $\theta_2$,
$\theta_A$ is the Abell radius and $P(\theta_1, \theta_2)$ is the flux
expected in an annulus around a point source in the PSPC. The two
denominators are normalising terms which remove the dependence on the
Abell richness and X-ray flux.

\begin{figure}
 \epsfxsize 0.95\hsize
 \epsffile{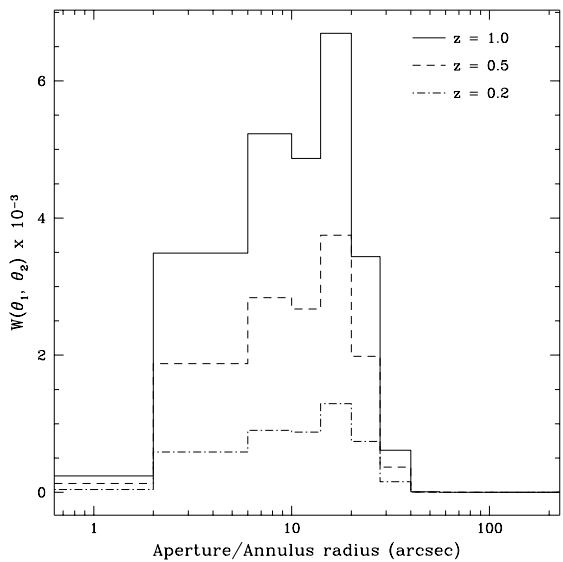}
 \caption{
Estimated correlation signal from galaxy clusters at a selection of redshifts.
}\label{fig:clustprof}
\end{figure}

The observed galaxy density distribution of clusters is well described
by a King model \cite{Sarazin86}:
\begin{equation}
\sigma(r) =
    \sigma_0 \left [ 1 + { {r^2} \over {r_c^2} } \right ]^{-1}
\label{eqn:king}\end{equation}
where $\sigma(r)$ is the projected density of galaxies at radius $r$ in Mpc,
$\sigma_0$ is the central density of galaxies and $r_c$ is the core radius of
the cluster in Mpc. We adopt a value $r_c = 0.25\,{\rm h}_{50}^{-1}$ Mpc
\cite{Bahcall75}. Evaluating eqn.~\ref{eqn:clustprof} for the on-axis PSPC
point spread function for clusters at a range of redshifts (H$_0$=50) we obtain
the results in figure~\ref{fig:clustprof}.
The form of $W(\theta_1, \theta_2)$ is largely independent of redshift
and is very similar to that seen for the ``bright'' galaxy/X-ray
correlation (figure~\ref{fig:18-22xcor}). This may indicate that a
significant fraction of this correlation signal is due to X-ray
emission from within clusters or groups of galaxies. The same signal
is not seen for the ``faint'' galaxy/X-ray correlation. This may be
due to a dilution of the signal from foreground and background
galaxies at these magnitudes. However, we would not necessarily expect
to see the same distribution for the ``faint'' sample since, although
there is no strong dependence on redshift, the King model given in
equation~\ref{eqn:king} is a good approximation only for the
distribution of cluster galaxies with magnitudes ${\rm m} < ({\rm m}_3
+ 2)$ where ${\rm m}_3$ is the third ranked cluster member. However,
the ``faint'' sample covers a range of only 1 magnitude, and so the
King model is not applicable.

\section{Conclusions}\label{sec:conclusions}

A significant correlation signal is seen between the distribution of
photons in the unresolved XRB and the positions of faint
galaxies. However, it is impossible to reliably determine the source
of the X-rays, with NELGs, hot intra-cluster gas and QSOs within
clusters all likely candidates. The angular form of the correlation
signal for the brighter galaxies is very similar to that expected for
emission from clusters over a range of redshifts, but the same signal
would clearly be seen for QSOs or NELGs at the centres of
clusters. Nevertheless, the increasing importance of NELG sources at
the fainter end of optically identified X-ray surveys and the
extrapolation of the observed QSO source counts to fainter fluxes both
imply that a significant fraction of the signal should come from
sources other than QSOs.

Comparison with simulations indicates that the correlation signal is
enhanced by clustering of galaxies. Although the angular form of this
signal is consistent with emission from the hot gas in moderately
distant clusters of galaxies, the angular scales are comparable to
that of the point spread function of the instrument, so no firm
conclusions can be drawn. Also, from the current data, we cannot
distinguish between X-ray emission from an intra-cluster medium and
emission from individual X-ray objects associated with clustered
environments. It is also important to remember that these results are
drawn from a single X-ray observation. Although the magnitude of the
observed signal is comparible to that of other, less-deep observations
(eg \pcite{Almaini+97}), the angular form of the signal is different.

However, it is clear from these results that the unresolved XRB {\em
beyond\/} the resolution limit of the faintest X-ray surveys has a
significant contribution from faint X-ray galaxies.

\end{document}